# A mathematical aid decision tool for RT planning

O. Sotolongo-Grau[1], D. Rodríguez-Pérez[1], J. A. Santos-Miranda[2], M. M. Desco[1], O. Sotolongo-Costa[3] and J. C. Antoranz[1,3]

[1] UNED, Departamento de Física Matemática y de Fluidos, Madrid, Spain
[2] Hospital General Universitario Gregorio Marañón/Oncology Service, Madrid, Spain
[3] Cátedra de sistemas complejos "Henri Poincaré", Universidad de la Habana, Cuba

*Abstract*— **It is possible to find the optimized radiation dose per session for a radiotherapy (RT) treatment, using a population dynamics model. This has already been done in a previous work for a protocol with 30 sessions and a fixed dose per session. Extending this model to other protocols, with a variable number of sessions, we could change the radiation dosage while keeping the success probability of treatment at its maximum value. This could help the RT oncology service managers to plan the sequence of patients and treatments adapting it to the facilities of the oncology service. Besides, if tumor surrounding tissue is not able to afford a high dosage, it could be useful to extend the treatment to a higher number of low dose radiation sessions, keeping an optimal treatment.**

## I. INTRODUCTION

There is an increasing concern about finding the suitable planning that maximizes the outcome of a radiotherapy (RT) treatment [1]. Even when prescribed dose can be applied in the programmed time [2], there are factors like the equipment technical maintenance stops that need to be considered [3]. On the other hand, waiting times have been shown to be a major problem in the achievement of high treatment efficiency [4, 5]. A tool to optimize the radiotherapy resources could be helpful to reduce the waiting times keeping therapy outcomes constant.

Recently, a mathematical model has been proposed to evaluate the efficiency of radiotherapy as a function of the dose and the tumor characteristics [6]. Although that model involves some parameters whose values are not precisely determined, it gives qualitative results in good agreement with clinical practice. Looking for an applicable although general method, RT treatment is modeled making the simplest possible assumptions. The existence of an optimum dosage which maximizes the treatment results was found there.

In the present work we show how this optimum dosage can be adapted to other protocols with a different number of sessions, while keeping the optimal probability of the "6 weeks, 5 sessions per week" reference protocol. This could be useful to oncology services to plan the RT treatments without decreasing treatment performance. With this the oncologists could adjust radiation dosage in order to minimize the damage being caused to the surrounding tissue.

## II. MATERIAL AND METHODS.

### A. Model

We will use a Lotka-Volterra like model to describe the tumor evolution grounded on some assumptions. Tumor cells growth $\dot{X}$ (as usual, a dot over a quantity represents its time derivative) depends on the current tumor population as $aX$ and its mass-law interaction with lymphocytes, $-bX$. Lymphocyte population grows due to tumor-immune system interaction, $dXY$, and contribution to exponential decay, $-fY$, due to natural cell death. The tumor is assumed to secrete interleukin which produces an immunity depression effect [7, 8], proportional to the number of cells in the tumor, $-kX$. In this model a constant flow, $u$, of lymphocytes arrives from the immune system.

So, we model tumor-immune system interaction using the already known equations [9]:

$$\dot{X} = aX - bXY$$
$$\dot{Y} = dXY - fY - kX + u \qquad (1)$$

Radiotherapy treatments are included in this equation using the LQ Model [10] for each radiation session. This brings a transformation of system (1) including a new equation for the tumor non clonogenic cells originated from radiation damage:

$$\dot{X} = aX - bXY - \dot{B}_t(t)X$$
$$\dot{Y} = dXY + pZY - fY - k(X+Z) + u - \dot{B}_l(t)Y \qquad (2)$$
$$\dot{Z} = \dot{B}_t(t)X - rZ - qZY$$

where $\dot{B}_{t,l}(T) = B_{t,l}\sum \delta(t - T_n)$ represent the amount of tumor cells and lymphocytes affected by radiation per unit time. $T_n$ are the time instants when radiation dose is applied and $\delta(t - T_n)$ denotes Dirac's delta function centered at $T_n$.

A dimensionless system can be easily obtained taking $t_c = 1/a$ (in absence of external influences) as the unit of time, $\tau = t/t_c$, $X = ax/d$, $Y = ay/b$ and $Z = az/d$:

$$\dot{x} = x - xy - \gamma_t(\tau)x$$
$$\dot{y} = xy + \varepsilon zy - \lambda y - \kappa(x+z) + \sigma - \gamma_l(\tau)y \qquad (3)$$
$$\dot{z} = \gamma_t(\tau)x - \rho z - \eta zy$$

where $\gamma_l(\tau)$, $\gamma_t(\tau)$ represent the dimensionless amount of lymphocytes and tumor cells affected by radiation per dimensionless unit time and, $\varepsilon = p/d$, $\lambda = f/a$, $\kappa = kb/ad$, $\sigma = ub/a^2$, $\rho = ra/d$ and $\eta = qa^2/db$. All the parameters can be estimated and interpreted as described in [11].

The parameter space analysis exhibits regions with different behaviors for the tumor and immune system interaction [6, 9]. The most relevant parameters are $\sigma/\lambda$, interpreted here as the efficiency of immune system over tumor growth, and $\kappa$, as the deficiency of immune system due to tumor growth. The appropriate region for radiotherapy treatments is that where $\sigma/\lambda < \kappa < 1$. In it, although the tumor grows exponentially, the immunity depression effects are moderate, so that patients maintain a good Karnofsky performance scale index [12] and fulfill those physical requirements to be subjects to treatment.

*B. Treatment optimization*

A reference protocol was mimicked in [6] while coefficients entering (3) were varied at random among admissible values [11]. The simulation, then, covers a wide range of tumors and supply general results useful to clinical practice. One million of "virtual patients" under treatment were simulated. Defining the probability of treatment success ($P_s$) as the fraction of "virtual patients" with no tumor at the end of treatment, it could be represented as a function of the tissue effect of the LQ Model $E(\alpha,\beta)$ (see [10]), and the Immune System Tumor Efficiency Rate [6], $ISTER$, defined as $ISTER = \sigma/\lambda$.

As shown in [6], to each value of $ISTER$ and $E$ corresponds a single value $P_s$. This means that the long term survival of patients will not improve with higher doses of radiation, on the contrary, it is possible to get the maximum success probability for our reference protocol at doses between 2 or 3 Gy [6]. Our model got that higher doses do not improve therapy outcomes.

*C. Simulation*

In this work, our simulation is basically the same as in [6] but differs in the fact that the number of sessions is not fixed. We mimic different radiation treatments with Eqs. (3) to simulate tumor evolution. To follow radiotherapy treatment in a realistic way, we apply a reference protocol with weekend interruptions. Furthermore, the protocols had variable length *i.e.*, are not restricted to 6 weeks. Each virtual patient is simulated with a variable number of sessions ($N$), from 15 to 40, beginning on the 10th day [1, 13].

To reproduce tumor evolution resembling a clinical case, one needs to calculate the correct values of the coefficients appearing in Eqs. (2). Numerical estimation of these coefficients was already made in [11] (and also in [15] for a slightly different model), based on clinically available data, showing a possible procedure for clinical professionals to estimate these values.

A statistical study of the dependence of treatment success on the dosage and number of radiation sessions has been performed. Due to the wide range of possible parameter values in Eqs. (3), their values have been drawn randomly from a log-normal distribution, to avoid negative values, but keeping the immune system efficiency ($\sigma/\lambda$) always smaller than one. The amount of cells affected by radiation, including lymphocytes and tumor cells, or the equivalent survival factors [10, 16], was taken as random values as in [6]. As initial conditions we assumed, for simplicity, the number of tumor cells is higher than the number of lymphocytes and both populations are distributed as normal random numbers. One million of virtual patients was generated taking different parameter values in Eqs. (3) and a fourth order Runge-Kutta method [14] was used to integrate the equations.

### III. RESULTS AND DISCUSION

The treatment success probability, $P_s$, was calculated as a function of $ISTER$, $N$ and $E$.

$$P_s = G(ISTER, N, E) \qquad (4)$$

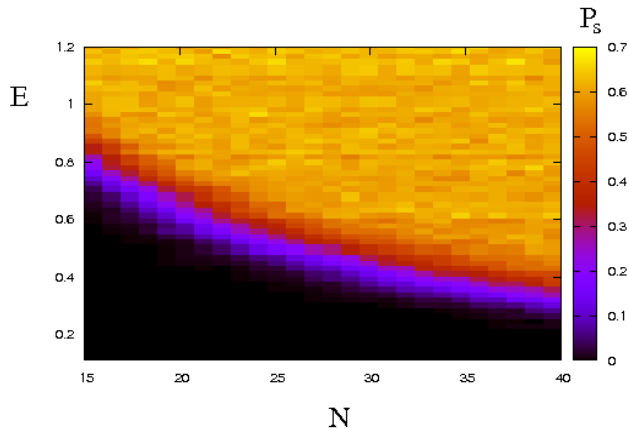

Fig. 1: $P_s$ representation for $ISTER = 0.7$ as a function of tissue effect and number of radiation sessions. Closer to black means lower, yellow means closer to 0.7.

As expected, for each value of $N$ we obtained similar results as those reported in [6]. Furthermore, when a fixed value of the *ISTER* is taken, the surface represented in figure 1 is obtained. This shows that whenever the value of $N$ increases, the optimized value of $P_s$ can be obtained with a lower value of $E$ per session. This allows us to find the exact values of $E$ that optimize $P_s$, for each value of $N$, as represented in figure 2.

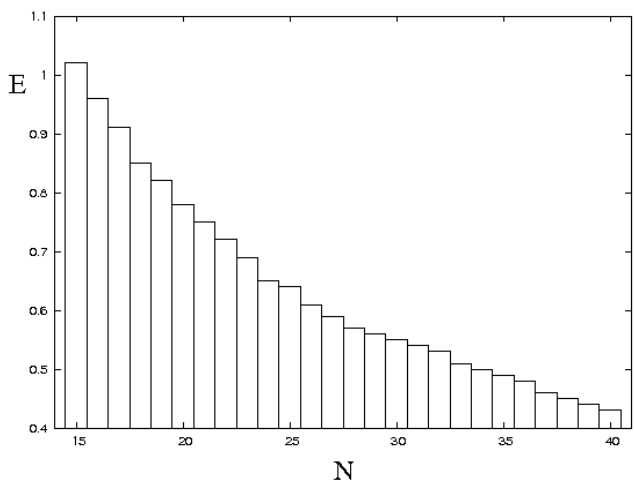

Figure 2: Tissue effect per session that maximize the success probability $P_s$ for each value of $N$. $ISTER = 0.7$.

For a tumor with known values of $\alpha$ and $\beta$ it is easy to find the amount of physical radiation per session ($d$) matching some $E$ value and hence the corresponding total amount of physical radiation $D$. To find the right combination of $N$ and $E$, the surrounding tissue properties facing radiation damage must be taken into account [10].

To illustrate the relevance of these results, let us consider an hypothetic patient with $\sigma/\lambda = 0.7$ and a tumor with $\alpha = 0.1 Gy^{-1}$ and $\alpha/\beta = 1 Gy$. Applying the 2Gy regular treatment dosage, each session has a tissue effect of $E = 0.6$. With this value of $E$, the treatment reaches its maximum success probability after 26 sessions. Subsequent sessions do not contribute to a better healing.

On the other hand, if under the same conditions, $\alpha/\beta = 5 Gy$, the optimal treatment involves 3.3Gy per session. If an oncologist wishes to radiate less than 3Gy per session, while keeping the optimal treatment, the treatment must be extended, at least, until the 36th session.

## IV. CONCLUSIONS

The present work presents a generalization of [6] to a scenario with a non standard number of sessions. The simulation allows us to find the corresponding tissue effect per radiation session providing the maximum value of the success probability. If, for instance, due to casual interruptions the RT machine could not be available for some time, our results could guide the radiotherapists to design a parallel treatment as efficient as that initially recommended for patients and adapted to the available time window.

Nonetheless, the reported results are not a definitive answer for the RT planning, since the surrounding tissue has not been taken into account here. If the surrounding tissue could not afford a high radiation amount per session, it could be also helpful to find an alternative treatment involving lower dosage whereas keeping the optimized success probability. The oncologist must then evaluate if a lower number of higher doses or higher number of lower doses would be less aggressive to the surrounding tissue and choose the appropriate one. Our results would help to improve the clinical results by assessing the amounts of radiation with better success probability.

## REFERENCES

1. Khoo, VS. (2005) Radiotherapeutic techniques for prostate cancer, dose escalation and brachytherapy. Clinical Oncology, 17:560-571.
2. González San Segundo C, Calvo Manuel FA, Santos Miranda JA. Retrasos e interrupciones: la dificultad para irradiar en el tiempo ideal. Clin Transl Oncol, 7.
3. Do V, Gebski V, Barton MB (2000). The effect of waiting for radiotherapy for grade iii/iv gliomas. Radiotherapy and Oncology, 57:131-136.

4. Mackillop WJ. (2007) Killing time: The consequences of delays in radiotherapy. Radiotherapy and Oncology, 84:1-4.
5. Jensen AR, Nellemann HM, Overgaard J. (2007) Tumor progression in waiting time for radiotherapy in head and neck cancer. Radiotherapy and Oncology, 84:5-10.
6. Sotolongo-Grau O, Rodriguez Perez D, et al. (2009) Immune system - tumor efficiency ratio as a new oncological index for radiotherapy treatment optimization. Math. Med. Biol., (Accepted for publication).
7. Whiteside TL. (2002) Apoptosis of immune cells in the tumor micro-environment and peripheral circulation of patients with cancer: implications for immunotherapy. Vaccine, 20:A46-A51.
8. Whiteside TL. (2006) Immune suppression in cancer: Effects on immune cells, mechanisms and future therapeutic intervention. Seminars in Cancer Biology, 16:3-15.
9. Sotolongo-Costa O, et al. (2003) Behavior of tumors under nonstationary therapy. Physica D, 178:242-253.
10. Steel GG. (1993) Basic Clinical Radiobiology for Radiation Oncologists. Edward Arnold Publishers, London.
11. Rodriguez-Perez D, Sotolongo-Grau O, et al. (2007) Assesment of cancer immunotherapy aoutcome in terms of the immune response time features. Math Med Biol, 24:287-300.
12. Sundstrom S, Bremnes R, et al. (2004) Hypofractioned palliative radiotherapy (17 Gy per two fractions) in advanced non-small-cell lung carcinoma is comparable to standard fractionation for symptom control and survival: A national phase III trial. Journal of Clinical Oncology, 22:801-810.
13. Rades D, Lang S. (2006) Prognostic value of haemoglobin levels during concurrent radio-chemotherapy in the treatment of oesophageal cancer. Clinical Oncology, 18:139-144.
14. Press WH, Teukolsky SA, et al. (1992) Numerical Recipes in C, The Art of Scientific Computing. Cambrige University Press, Cambrige.
15. de Pillis L, Radunskaya AE, Wiseman CL. (2005) A validated mathematical model of cell-mediated immune response to tumor growth. Cancer Research, 65:7950-7958.
16. Enderling H, Alexander RA, Mark AJ. (2006) Mathematical modelling of radiotherapy strategies for early breast cancer. Journal of Theoretical Biology, 241:158-171.